\documentclass[conference]{IEEEtran}
\ifCLASSINFOpdf
\else
\fi
\usepackage{verbatim}
\usepackage{cite}
\usepackage{fancyhdr}
\usepackage{algorithm}
\usepackage{algorithmic}
\usepackage[small]{caption}
\usepackage{graphics}
\usepackage{color}
\usepackage{subfigure}
\usepackage{slashbox}
\usepackage[hyphens]{url}
\usepackage{hyperref}
\usepackage{epstopdf}
\usepackage{multirow}
\usepackage{underscore}
\usepackage[dvips]{graphicx}
\usepackage{array}

\begin{document}
%
\title{bit.ly/malicious: Deep Dive into Short URL based e-Crime Detection}

\author{
\IEEEauthorblockN{Neha Gupta, Anupama Aggarwal, Ponnurangam Kumaraguru }
\IEEEauthorblockA{Indraprastha Institute of Information Technology, Delhi (IIIT-D)}
\IEEEauthorblockA{Cybersecurity Education and Research Centre, IIIT-Delhi }
\IEEEauthorblockA{\{neha1209, anupamaa, pk\}@iiitd.ac.in}
}

%


\maketitle
\thispagestyle{empty}
\pagestyle{empty}
\begin{abstract}
Existence of spam URLs over emails and Online Social Media (OSM) has become a massive e-crime. To counter the dissemination of long complex URLs in emails and character limit imposed on various OSM (like Twitter), the concept of URL shortening has gained a lot of traction. URL shorteners take as input a long URL and output a short URL with the same landing page (as in the long URL) in return. With their immense popularity over time, URL shorteners have become a prime target for the attackers giving them an advantage to conceal malicious content. Bitly, a leading service among all shortening services is being exploited heavily to carry out phishing attacks, work-from-home scams, pornographic content propagation, etc. This imposes additional performance pressure on Bitly and other URL shorteners to be able to detect and take a timely action against the illegitimate content. In this study, we analyzed a dataset of 763,160 short URLs marked suspicious by Bitly in the month of October 2013. Our results reveal that Bitly is not using its claimed spam detection services very effectively.
 We also show how a suspicious Bitly account goes unnoticed despite of a prolonged recurrent illegitimate activity. Bitly displays a warning page on identification of suspicious links, but we observed this approach to be weak in controlling the overall propagation of spam. We also identified some short URL based features and coupled them with two domain specific features to classify a Bitly URL as malicious or benign and achieved an accuracy of 86.41\%. The feature set identified can be generalized to other URL shortening services as well. To the best of our knowledge, this is the first large scale study to highlight the issues with the implementation of Bitly's spam detection policies and proposing suitable countermeasures.~\footnote{We thank Bitly for sharing the data with us. In particular, we have interacted with Brian Eoff, Lead Data Scientist at Bitly by sharing our analysis and getting his reactions to our conclusions.}
\end{abstract}


%
\IEEEpeerreviewmaketitle

\section{Introduction}
\vspace{4pt}
URL shortening is a technique of mapping a long Uniform Resource Locator (URL) to a short URL redirecting to the same landing page. Initially, the concept was used to prevent breaking of complex URLs while copying text, to accommodate long URLs without line breaks, and for smooth dissemination of content. Usage of these services nowadays has become a trend in Online Social Media (OSM); content length restriction imposed by various OSMs (e.g. Twitter's 140 character limit) has helped popularize their use even further. In order to accommodate more content in their tweet, users prefer to compress their long URLs using URL shortening services. Some popular URL shortening services like \textit{bit.ly} and \textit{goo.gl} track URLs and provide real time click traffic analysis \cite{25,29}. Although these services are created to comfort the users, spammers have found their ways to target and misuse the facility for their benefit.

URL shorteners do not only reduce the URL length but also obfuscate the actual URL behind a shortened link. Spammers take advantage of this obfuscation to misguide the netizens by posting malicious links on OSM. These malicious links can be: i) \textit{spam} - irrelevant messages sent to large number of people online, ii) \textit{scam} - online fraud to mislead people, iii) \textit{phishing} - online fraud to get user credentials, or iv) \textit{malware} - auto downloadable content to damage the system.~\footnote{This is not a comprehensive list and there can be other type of illegitimate content that we did not mention here.} Link obfuscation makes short URL spam more difficult to detect than traditional long URL spam. Malicious long URLs can be detected with a direct domain lookup or a simple blacklist check, while short URLs can easily escape such technique. Lexical methods to detect long URL spam work slow for short URLs because of additional redirects. Malicious short URL detection therefore requires more efficient spam detection techniques. According to a threat activity report by Symantec in year 2010 \cite{20}, around 65\% malicious URLs on OSM were shortened URLs. Another study in 2012 investigated a particular URL shortening service (\textit{qr.cx}) and revealed that around 80\% of shortened URLs from this service contained spam-related content \cite{2}. Research by a URL shortener \textit{yi.tl} reveals that because of deep penetration of spam, 614 out of 1,002 URL shortening services became non-functional in year 2012.~\footnote{\url{http://www.webpronews.com/study-claims-61-of-url-shorteners-are-dead-2012-05}} A recent article highlights that Facebook spammers make close to 200 million dollar through posting these shortened links to lure users \cite{21}.

Bitly, launched in year 2008 is one of the most popular URL shortening services on the web \cite{18}. It gained major traction when Twitter started to use it as a default URL shortener in year 2009 before the launch of its own service, \textit{t.co} in the year 2011.~\footnote{\url{https://support.twitter.com/articles/109623-about-twitter-s-link-service-http-t-co}} Bitly provides an interface to its users to either shorten a link anonymously or create an account to shorten the links. Each link shortened by a user has a unique \textit{global hash} (an aggregated identifier corresponding to a link). Such shortened links, known as \textit{bitmarks} can then be saved, tracked, and shared. Users are also allowed to connect any number of Facebook / Twitter accounts with their Bitly accounts, making the task of shortening and sharing a link very convenient. With 1 billion new links shortened on Bitly each day and 6 billion clicks each month, spammers have been exploiting the service to a great extent \cite{27}. In early 2013, a news article reported the spread of phishing attacks on Twitter through Direct Messages (DM) with malicious Bitly links.~\footnote{\url{http://news.softpedia.com/news/Twitter-Phishing-Scam-This-Profile-Is-Spreading-Nasty-Blogs-Around-About-You-318618.shtml}} Large number of users fell in the trap and clicked on the link, which redirected to a website that replicated Twitter's login page. Victims were then misled to believe that their session was expired and were made to login again, unknowingly revealing their Twitter credentials to the attacker. Impact of the attack was such that Twitter announced a temporary restriction on sending shortened links including Bitly in DMs.~\footnote{\url{http://www.digitaltrends.com/social-media/twitter-may-be-banning-links-in-dms/\#!CvRAI}}

In another attack, spammers abused the redirect vulnerabilities of a popular legitimate domain belonging to the U.S. federal government, which had collaboration with Bitly. The hijacked domain \textit{1.usa.gov} which redirected to an illegitimate \textit{work-from-home} scam website received around 43,049 clicks from 124 countries within a week.~\footnote{\url{http://www.pcworld.com/article/2012800/spammers-abuse-gov-url-shortener-service-in-workathome-scams.html}} This shows that even \textit{branded short domains} by Bitly are not safe from exploits \cite{28}. 
In October 2013, Bitly also experienced a massive DDoS attack rendering complete shutdown of its services for close to 7 hours.~\footnote{\url{http://www.geeknewscentral.com/2013/10/21/twitter-banning-bit-ly-other-url-shortners-on-direct-messages-dm/}} Some spammers have started to build their own URL shortening services to double-shorten the malicious links, first with a self created short URL service, then with a legitimate short URL service to evade security checks.~\footnote{\url{http://www.geek.com/news/fake-short-url-services-latest-tactic-for-spammers-1382555/}} Security researchers from Symantec found that spammers used Bitly URLs to propagate sexually suggestive content \cite{16}. 

Unlike other URL shortening services like \textit{goo.gl} and \textit{ow.ly}, Bitly does not provide a CAPTCHA to test human identity at the time of URL shortening. For protection against spam, Bitly claims to use real-time spam detection services like Google safebrowsing and SURBL, and flags 2-3 millions links as spam each week \cite{23,24,26}. Bitly neither deletes a flagged suspicious link nor suspends the associated user; but displays a warning page whenever such a link is clicked. Such a warning page is by-passable and does not completely restrict a user to visit the malicious website. Also, non-deletion of illegitimate content or account can make it viral over web. Despite of all these detection measures adopted by Bitly, there is continued existence of malicious Bitly URLs. It is therefore important to have an in depth understanding of the gaps in Bitly's spam detection techniques that deter its efficiency to handle malicious content. This paper deals with the identification of such gaps and highlights some countermeasures which can be adopted by Bitly to be able to detect malicious content more effectively.

In this work, we perform a detailed analysis on a dataset of suspicious Bitly links and their associated attributes to characterize Bitly spam and explore its spam detection policies. Major contributions of this paper are:
\begin{itemize}[leftmargin=0.4cm]
\item \textit{Impact analysis of malicious Bitly links on OSM}: There exist large communities propagating malicious content through Bitly on Twitter. Such communities can grow in size if Bitly does not impose any limit on the number of connected OSM accounts. 
\item \textit{Identification of issues in Bitly's spam detection}: We found that Bitly is unable to detect malicious links tracked by popular blacklist services and is not using its claimed spam detection techniques very effectively. Spammers exploit Bitly's \textit{no account suspension} policy and keep shortening malicious URLs. 
\item \textit{Machine learning classification to detect malicious Bitly URLs}: Our classification mechanism relies on the combination of long and short URL based features and we attained an accuracy of 86.41\%. Our technique can work efficiently irrespective of the number of clicks received by a Bitly URL.
\end{itemize}
To the best our knowledge, this is the first large scale study to highlight the issues with Bitly's spam detection policies and propose a suitable solution.
The remainder of the paper is organized as follows: Section 2 presents the related work and Section 3 explains our data collection methodology. Analysis and results are covered in Section 4 and 5. Section 6 contains the conclusive summary and Section 7 presents some future directions and limitations of our research.

\section{Related Work}
\vspace{4pt}
URL shortening services take as input a long URL (e.g.~\textit{\url{http://blog.bitly.com/post/138381844/spam-and-malware-protection}}) and generates a short URL (e.g.~\textit{\url{bit.ly/RSwVGo}}) in return. Short URL so generated redirects to the same long URL but looks random and unrelated to the actual link. Imposed character limit has lead to immense popularity of such services in social media landscape. Due to their ubiquitous usage, these services have been hit by adversaries to obfuscate and disseminate malicious content. This section presents the work done in understanding the usage pattern, behavior, and misuse of short URLs.
\subsection{Malicious Long URL Characterization / Detection}
\vspace{4pt}
A number of studies have been conducted to understand the propagation of spam on OSM, many of which revealed heavy usage of URLs to spread malicious content. Benevenuto et al. in their research identified distinctive features to detect spammers on Twitter \cite{8}. Researchers also evaluated the effectiveness of popular blacklists in evading spam and observed it to be inefficient. On Twitter, checking blacklists becomes even slower because of the URL shortening services used to obfuscate long URLs. Using these services, a spammer can complicate the process of detection by using chains of multiple shortenings~\cite{10}. 
 Thomas et al. in 2011 developed a system called Monarch which classifies a URL submitted to any web service as malicious or benign in real time \cite{6}. This system relies on the features of URL landing page (like page content, hosting infrastructure, etc.) and detected malicious links with an accuracy close to 91\%. In year 2012, Aggarwal et al. also proposed a real time phishing detection system for Twitter, called PhishAri \cite{9}. Authors coupled the Twitter and URL based features to classify phishing tweets and achieved an accuracy of 92.52\%. 
In another real time suspicious URL detection technique on Twitter proposed by Lee et al., authors addressed the problem of conditional URL redirects \cite{11}. A combined feature set of \textit{correlated URL redirects} and \textit{tweet context information} was used and authors attained an accuracy of 86.3\%. 
\subsection{Short URL Analysis}
\vspace{4pt}
With the introduction of short URLs in OSM, a comparative study is necessary to be able to understand the level of acceptance of short URLs over long URLs. Kandylas et al. performed a comparative study of long and short Bitly URLs propagation on Twitter and found that Bitly links received orders of magnitude more clicks than an equal random set of long URLs \cite{5}.  
To further comprehend short URL distinctive characteristics, Antoniades et al. studied the lifetime of short URLs which revealed that the life span of 50\% short URLs exceeds 100 days. 
Other than this generic analysis, Neumann et al. looked at malicious short URLs in emails and highlighted their privacy and security implications \cite{1}. Chhabra et al. also gave an overview of evolving phishing attacks through short URLs on Twitter and found that phishers use URL shorteners not only to gain space but hide their malicious links \cite{13}. Their results show that most of the tweets containing phishing URLs comes from inorganic (automated) accounts. Later in year 2012, Klien et al. presented the global usage pattern of short URLs by setting up their own URL shortening service and found 80\% short URL content to be spam related \cite{2}. 
In year 2013, Maggi et al. performed a large scale study on 25 million short URLs belonging to 622 distinct URL shortening services \cite{4}. Their results highlight that the countermeasures adopted by these services to detect spam are not very effective and can be easily by-passed. Experimental results from their data shows that Bitly allows users to shorten malicious links and does not include any initial level lightweight check to prevent it (though detects it after some time). Unlike their study which focused on multiple URL shorteners, our research is an in depth analysis of the effectiveness of a single URL shortener. Another scheme was proposed by Yoon et al. in year 2013 about using relative words of target URLs in short URLs \cite{14}. This can give hints to user to guess the target URL, making it comparatively safe from phishing attacks.

\subsection{Malicious Short URL Characterization / Detection}
\vspace{4pt}
There is little research done in the area of malicious short URL characterization. One such work that presents short URL based features to detect malicious accounts is given by Wang et al. in year 2013 \cite{12}. In their experiment, they investigated the creators of 600,000 short Bitly URLs and associated click traffic from different countries and referrers. Based on the analysis, they classify a link as spam / non-spam using only 3 click traffic based features with maximum accuracy of 90.81\%, but ignored all short URLs with zero clicks. Our study on the other hand incorporates all URLs irrespective of their click state. In addition, their results reveal that legitimate Bitly users also generate spam and most clicks on short malicious URLs comes from popular websites. 

After reviewing the above literature, it is evident that a lot of work has been done in the identification and analysis of malicious URLs. Surprisingly, very little work has been done in analyzing only suspicious short URLs to expose the gaps in security mechanisms adopted by a specific URL shortener. Our work significantly differs from the prior studies, as we focus on understanding in depth, the loopholes in spam detection mechanisms of a URL shortening service. We also propose and evaluate a semi supervised classification framework for spam detection in URL shorteners. 
\section{Data Collection}
\vspace{4pt}
For our experimental dataset, we followed a two-phase approach. In the first phase, we acquired a dataset of suspicious Bitly URLs from Bitly and collected their associated attributes to explore the issues with Bitly's spam detection techniques. In the second phase, we collected a dataset of Bitly URLs from Twitter and used machine learning algorithms to classify an unknown Bitly URL as malicious or benign. We call a Bitly URL as benign if it is non-malicious and trustworthy.
\subsection{Data Collection Methodology (Phase 1)}
\vspace{4pt}
To analyze the basic characteristics of short URL spam, we requested Bitly to share with us the links that they mark as suspicious. We received a dataset of 763,160 suspicious Bitly URLs which displayed a Bitly warning page in the month of October 2013. This dataset comprised of the global hash, associated long URL, and number of warning pages displayed for the global hash. We call this the \textit{link-dataset}. Bitly also provides a public API~\footnote{\url{http://dev.bitly.com/api.html}} to extract the link and user metrics for a particular short URL. Using the \textit{link-dataset} as our seed input to Bitly API, we collected analytics for 144,851 (18.98\%) links between January 2014 to March 2014 (our data collection is still on). Table~\ref{BitlyAnalytics} presents information about these analytics. We call this the \textit{link-metric-dataset}.
\begin{table}[h]
\small
 \begin{center}
    \begin{tabular}{|p{2.5cm}|p{4cm}|}
    \hline
    {\bf Short URL metric}     & {\bf Output data}                                                             \\ \hline
    info                 & link creator and creation time                  \\ \hline
    expand               & target long URL                                                         \\ \hline
    clicks               & last 1,000 click history                                                 \\ \hline
    referring\_domains   & domains referring click traffic to the given Bitly URL                  \\ \hline
    encoders             & users who created the given Bitly URL                                        \\ \hline
    encoder\_info        & Bitly profile name, creation date, and connected networks \\ \hline
    encoder\_link\_history & last 100 links of the encoder                                 \\ \hline
    \end{tabular}
    \caption {\label{BitlyAnalytics} Data we obtained for the suspicious Bitly links.}
 \end{center}
\end{table}
\subsection{Data Collection Methodology (Phase 2)}
\vspace{4pt}
In order to collect an unlabeled dataset (mix of benign and malicious), we used Twitter Rest API~\footnote{\url{https://dev.twitter.com/docs/api/1.1}} and its ``search" method to get only tweets with a Bitly URL. Here we restricted our search query to ``bit.ly" and collected a total of 412,139 tweets with 34,802 distinct Bitly URLs between 12 February 2014 to 15 March 2014. We refer to this as the \textit{unlabeled-dataset} in the rest of this paper.
 To obtain a labeled dataset, we queried Google Safebrowsing, SURBL, PhishTank, and VirusTotal~\footnote{\url{https://www.virustotal.com/}} APIs to find whether the URLs from \textit{unlabeled-dataset} are malicious or not. Google Safebrowsing is a repository of suspected phishing or malware pages maintained by Google Inc. The Google Safebrowsing API accepts an HTTP GET / POST request to lookup a URL and returns a JSON object describing whether the URL is ``phishing'', ``malware'', or ``ok''.  SURBL is a consolidated list of websites that appear in unsolicited messages. SURBL lookup feature allows a user to check a domain name against the ones blacklisted by SURBL. We used SURBL client library implemented in python.~\footnote{\url{https://pypi.python.org/pypi/surblclient/}} PhishTank is a public crowdsourced database of phishing URLs where contributors submit suspicious URLs and volunteers label them as phishing or legitimate.~\footnote{\url{https://www.phishtank.com/}} The PhishTank API uses an HTTP post request and returns the status of a URL in standard JSON format. VirusTotal is an aggregated information warehouse of malicious links and domains as marked by 52 website scanning engines and contributed by users. The VirusTotal API also allows an HTTP POST request and gives a JSON response indicating results from all website scanning engines it uses. To label our dataset, we mark a Bitly URL as malicious if the expanded URL or domain is detected by any of these blacklists.  

\begin{figure}[h!]
 	\centering
		\includegraphics[scale = 0.34]{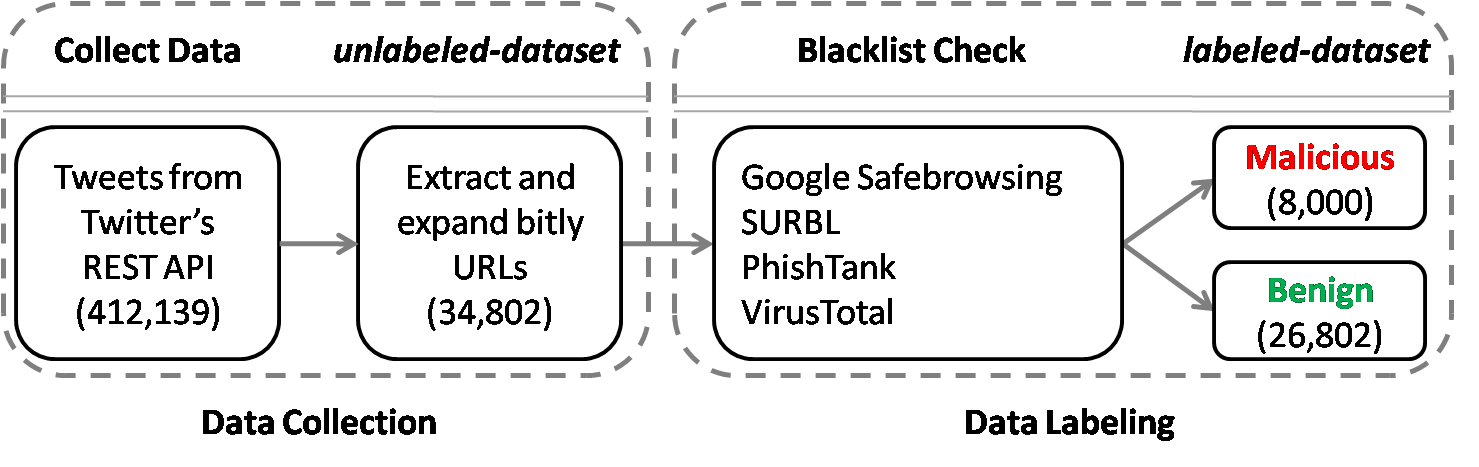}
	\caption{Data collection and labeling.}
		\label{dataCollectionLabeling}
\end{figure}

In addition, we also label a Bitly URL as malicious if it is detected by Bitly itself. Bitly uses various blacklisting services and other measures to detect spam and throws a warning page whenever it identifies a malicious URL. We perform this check for all Bitly URLs in our \textit{unlabeled-dataset} and label a URL as malicious if a warning page is displayed. Using these techniques, we obtained 8,000 distinct malicious Bitly URLs from our \textit{unlabeled-dataset} of 34,802 Bitly URLs. We call this the \textit{labeled-dataset}. Figure~\ref{dataCollectionLabeling} shows this data collection process.
\section{Analysis and Results}
\vspace{4pt}
In this section, we focus on the analysis of dataset obtained from Bitly. Our objective is to highlight some characteristics of malicious short URLs and to underline the weaknesses in security mechanisms used by Bitly. We attempt to answer some unexplored questions related to short URL spam like: (i) What are the characteristics of domains from where such malicious content is originating? (ii) Does malicious Bitly links have an impact on OSM? (iii) Is Bitly using the claimed spam detection services effectively? (iv) Do spammers take advantage of Bitly's \textit{no account suspension} policy? (v) Does a warning page alone help curtail the overall problem of spam? (vi) How quick is Bitly in identifying suspicious accounts?

Answering these questions is important to investigate Bitly's competence in dealing with illegitimate content. A detailed investigation is needed to understand the ground issues with Bitly's malicious content detection policies in order to make it more effective. 
\subsection{Domain Analysis}\label{Domain Analysis}
\vspace{4pt}
Quick and easy availability of domains has made the task of a spammer more convenient. Our \textit{link-dataset} of 763,160 suspicious URLs comprised of 22,038 unique domains. Since our target was to analyze only the malicious domains, we realized that the spammers often exploit some legitimate popular domains to propagate spam. Keeping this in mind, we used APWG (Anti-Phishing Working Group~\footnote{\url{http://www.antiphishing.org/}}) whitelist and separated all the legitimate domains from our \textit{link-dataset}. We found that 56 domains marked suspicious by Bitly were whitelisted by APWG.
Ignoring these, we created a python crawler and performed a test on the existence of each suspicious domain 5 months after we received our dataset. We observed that 83.06\% domains no longer existed. This highlights that such domains are actually short lived and created with a dedicated purpose of spamming. To further estimate the average number of times users tried to visit the links from such domains, we looked at the cumulative count of corresponding Bitly warning pages. Total number of click requests made to the URLs belonging to non-functional domains only in the month of October was found to be 9,937,250. Spammers thus focus on buying domains to host malicious content and most of these domains eventually die after achieving a good number of hits.
\subsection{Connected Network Impact Analysis}\label{Connected Network Impact Analysis}
\vspace{4pt}
Bitly allows its users to connect to any number of Facebook / Twitter accounts. This help users to shorten and share links at one click on the connected OSM. In this section, we present how Bitly users take advantage of this service for spamming. To investigate, we first extracted all the encoders~\footnote{We use `encoders' and `users' interchangeably.} of URLs in our \textit{link-metric-dataset} and found 12,344 distinct Bitly users from 413,119 malicious Bitly URLs. Next, we used the Bitly API to collect information about their connected social network and
 found 3,415 (63.54\%) users connected only Twitter, 951 (17.69\%) only Facebook, and 1,009 (18.77\%) connected both.
 Possible explanation for low Facebook connections could be that Bitly allows a user to connect a personal Facebook account for free, but linking Facebook brand or fan pages is a paid service.~\footnote{\url{http://support.bitly.com/knowledgebase/articles/76455-how-do-i-connect-my-facebook-and-twitter-account-t}} 
 This might restrict spammers to disseminate more malicious content in public, but there is no such documented limitation for Twitter. 

On further inspection, we found that 507 Bitly users connected multiple Twitter accounts of which 28 users connected at least 10 Twitter accounts. To analyze these 28 suspicious profiles, we extracted their last 200 tweets using Twitter Rest API. Twitter API gives last 200 tweets of a profile on a single request, and we believe it to be a reasonable sample for our study. Our target was to compare multiple profiles connected by each user and infer a possible reason behind their connection. We extracted the URLs posted by all Twitter accounts connected to one Bitly user and did a cross URL / domain comparison. For this we computed pair wise Jaccard similarity score~\footnote{\url{http://www.stanford.edu/~maureenh/quals/html/ml/node68.html}} followed by the overall variance. We also collected the link history (last 100) of these users using the Bitly API. All these links were checked for a Bitly warning page by making GET requests in python. At last, we manually inspected these accounts by looking at the tweet text and URL similarity scores. From this we identified 3 different cross-network communities that existed across Bitly and Twitter in our dataset:
\subsubsection{Community 1}The first community consisted of 27 Bitly users with 1 associated Twitter account each. All these users shortened links from the same domain and had similar looking user handles starting with ``o\_'' followed by some random string (e.g. o\_16ee0qg6i6). Last 100 links shortened by all these users redirected to a Bitly warning page. Also, all the associated Twitter accounts were suspended when we checked these profiles later and the Bitly profiles looked dormant. This behavior confirms the existence of this malicious community which used Bitly as a medium to propagate spam on Twitter. 
\subsubsection{Community 2}Another community comprised of 2 Bitly users with 28 associated Twitter accounts each. Also, during the course of our study one of these Bitly accounts connected 2 more Twitter accounts. Thirteen of the accounts did not exist when we rechecked and other 45 looked malicious and shared similar tweet text and URLs. This community appears to be conducting an active spam campaign. 
\subsubsection{Community 3}The third community composed of 2 Bitly accounts with 9 Twitter accounts each. All these 18 Twitter accounts shared similar explicit pornographic content. On a manual inspection of this Bitly profiles, their last activity was a year before and they were no longer shortening URLs. On the contrary, they were still posting tweets from their Twitter accounts. This shows that this malicious community is dormant on Bitly but still active on Twitter. 

These communities originated from Bitly to spread malicious content on Twitter. Presence of such big communities propagating malicious content clearly highlights the abuse of connected social network on Bitly. Bitly should therefore impose a restriction on the number of OSM accounts a user can connect.
\subsection{Analysis of Bitly's spam detection techniques}
\vspace{4pt}
Till now, we looked at a generalized characterization of malicious Bitly links by studying the domains they arrive from and the connected social network. The above results highlight that spammers use Bitly as a start point to propagate spam over other media. It now becomes important to comprehend whether Bitly is taking enough measures to deal with such content and the users it is coming from. In this section, we do a focused study to understand how Bitly reacts in this situation.
\subsubsection{Efficiency Analysis}\label{Efficiency Analysis}
To infer the efficiency of Bitly in detection of malicious links and users, we conduct a two-step experiment.

\vspace{4pt}
\textbf{Comparison With Popular Blacklists}

We performed a check for malicious Bitly links detected by 3 popular blacklist services- APWG, VirusTotal, and SURBL.
 To collect data from APWG, we requested for APWG's live feed service and set it up on MySql database. We collected the data for 6 months (October 2013 - March 2014) and obtained a total of 142,660 and a daily feed of around 746 APWG marked malicious links. On a direct lookup, we got 216 Bitly links from this dataset. In order to extract more Bitly links, we performed a reverse lookup by shortening the long malicious URLs and checking their existence on Bitly using Bitly API. Whenever a link is shortened using Bitly API, it returns a parameter called \textit{new\_hash}, indicating whether this link is shortened for the first time or is pre-existent. We collected only the pre-existent links. With the direct and reverse lookup, a total of 2,872 APWG marked malicious Bitly links were obtained. We also made GET requests to Bitly to check if it gives a warning page for these malicious URLs. To our surprise, Bitly could not detect 2,490 of 2,872 (86.70\%) malicious links. Though Bitly does not claim to use APWG, but such high non-detection rate looks alarming as APWG is a popular and trusted source to detect phishing. In addition to APWG, we also collected malicious links marked by VirusTotal over the same time period. For this, we implemented a web crawler and set up a cron job to perform daily look ups on VirusTotal and received 569 malicious Bitly links. We again checked these links for a Bitly warning page and found 407 of 569 (71.53\%) links undetected by Bitly. These results clearly highlights that Bitly misses on a lot of malicious content.

\begin{figure*}[ht]
 	\centering
		\includegraphics[scale = 0.68]{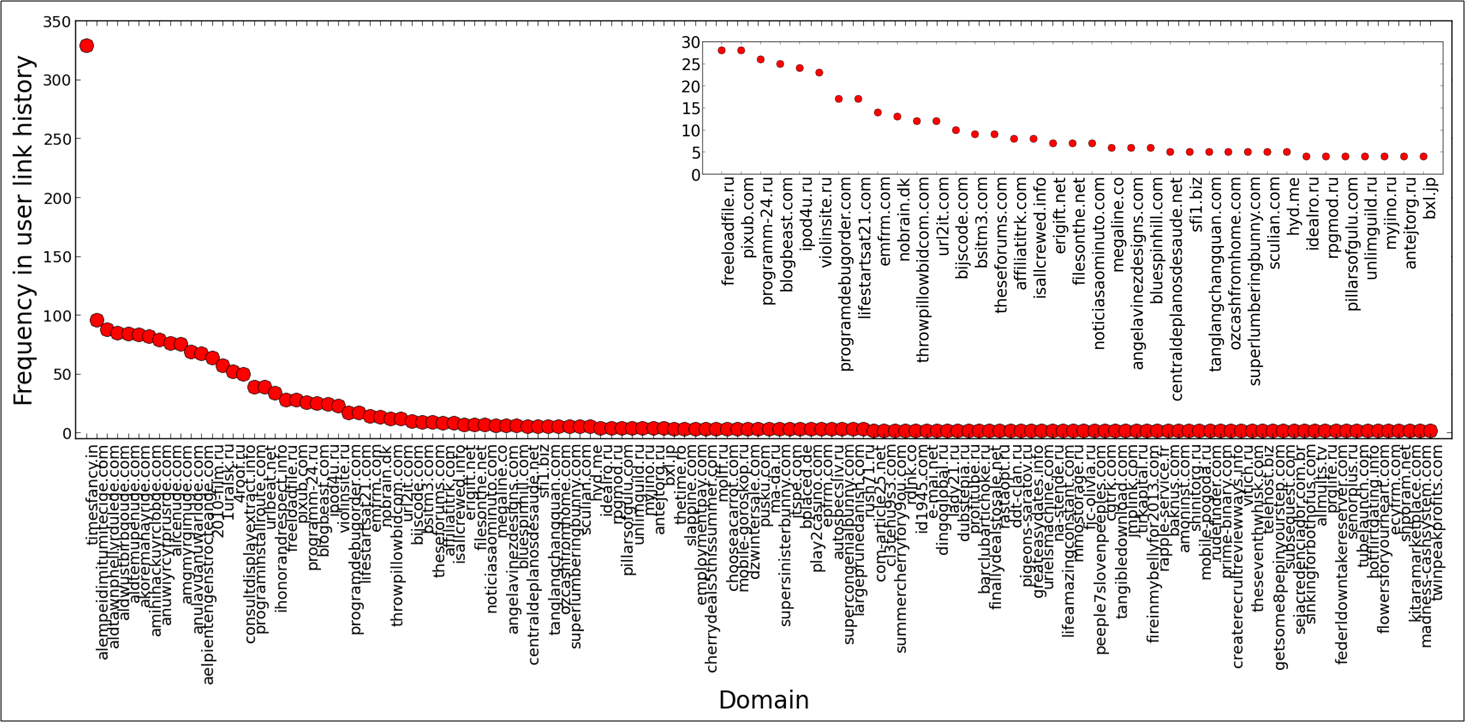}
	\caption{Frequency distribution of SURBL domains undetected by Bitly (with frequency more than 1). Blacklisted domain \textit{timesfancy.in} has the maximum frequency and 16 domains have frequency greater than 50. Inset: Starting from domain \textit{freeloadfile.ru} shows the frequency distribution within the graph more clearly.}
		\label{surblDomainsUndetected}
\end{figure*}

In addition to these popular blacklists which Bitly does not claim to use, we considered a third measure (SURBL) that Bitly professes to apply. 
 To check against SURBL, we used Bitly API to collect the link history (last 100 or less) of all encoders in our \textit{link-metric-dataset}. We received 717,644 Bitly links from 12,344 distinct Bitly encoders contributing to 63,693 distinct domains. On checking these links for a Bitly warning page and the corresponding domains against SURBL, we found 275 (36.66\%) domains blacklisted by SURBL but undetected by Bitly. 
These 275 domains contributed to 2,244 links in our dataset. Figure~\ref{surblDomainsUndetected} presents the frequency distribution of undetected domains with occurrence more than one. Inset in the same Figure starting from domain \textit{freeloadfile.ru} highlights the frequency distribution more clearly. This shows that there were multiple Bitly links corresponding to 129 of these domains, maximum being 329 for domain \textit{timesfancy.in}. Also, there were 16 domains with frequency more than 50. These results show that in addition to other popular blacklists, Bitly is also not using the claimed spam detection services very effectively. Such undetected domains contribute to a large number of links if looked at a greater scale. Thus, letting bypass a single malicious domain can act as an invitation to a huge amount of illegitimate content.
\begin{figure*}[ht]
\centering
\subfigure[]{
    \label{cumulative}
    \includegraphics[scale = 0.18]{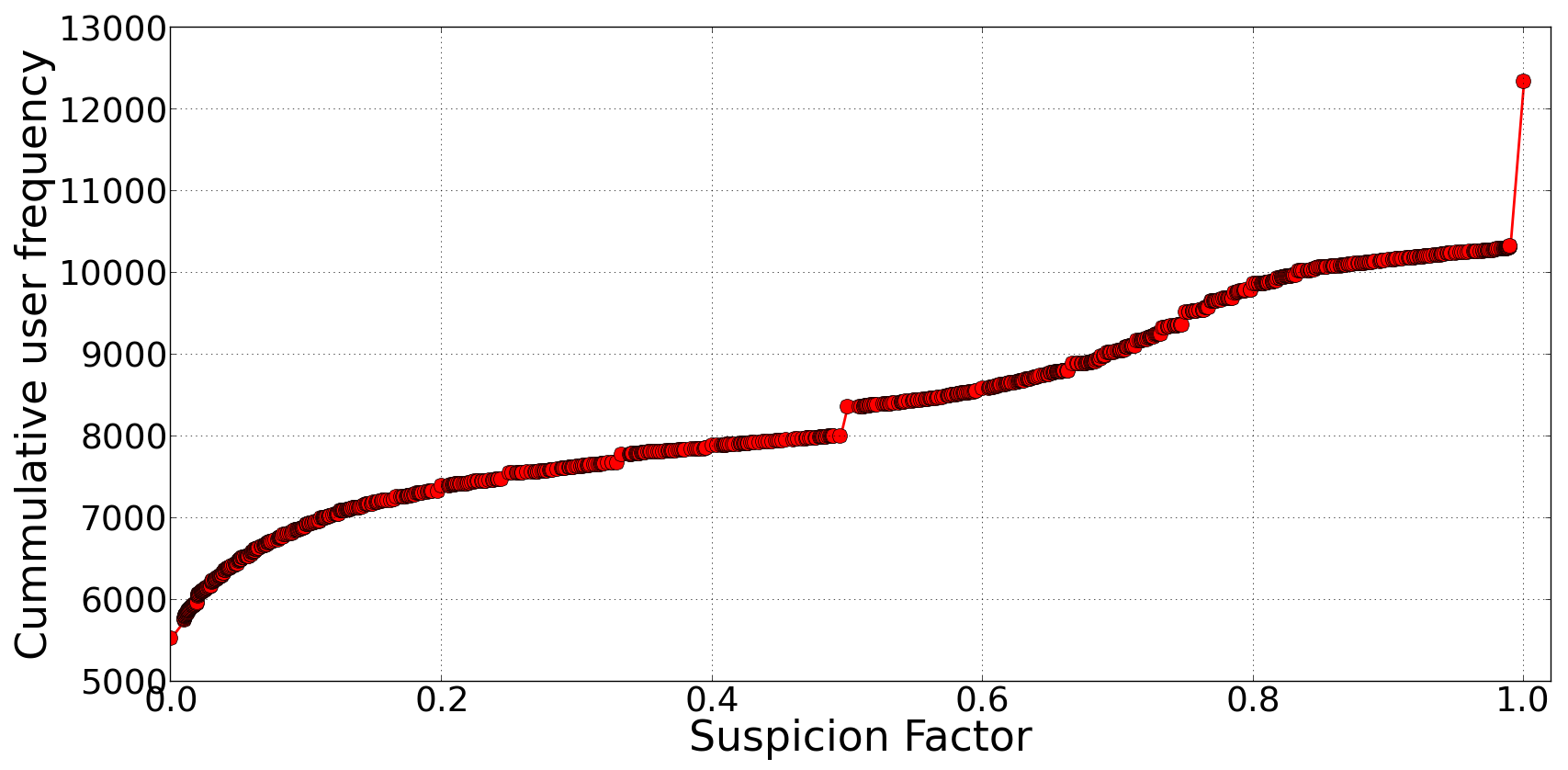}}
\subfigure[]{%
     \label{fig:bamsesang}
     \includegraphics[scale=0.18]{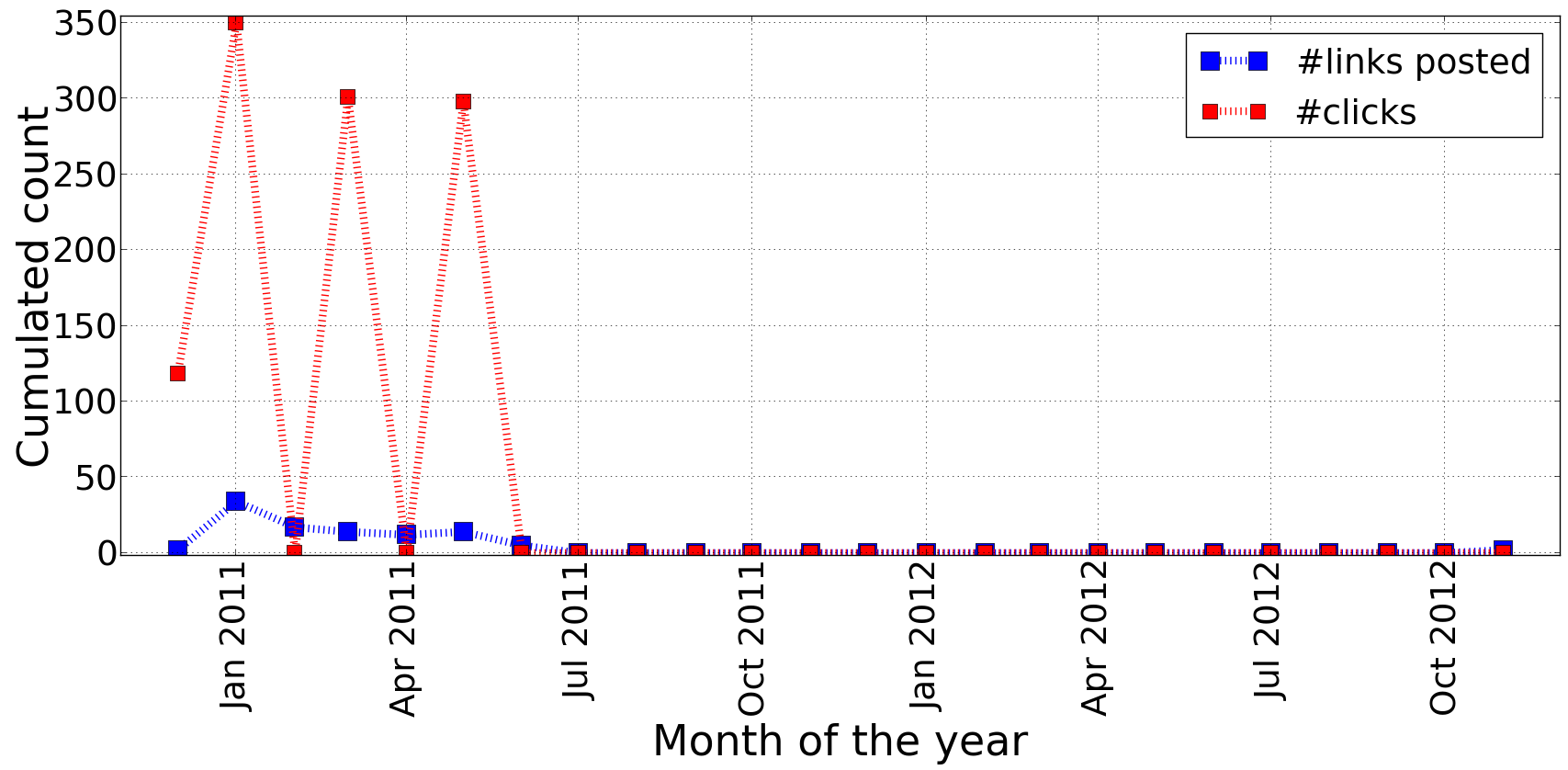}}
\caption{(a) Cumulative distribution on number of Bitly users posting suspicious links. (b) Link history timeline for user \textit{bamsesang}. The link sharing interval and click pattern clearly reflects the malicious activity being carried out for a long time.}
\end{figure*}

\vspace{4pt}
\textbf{Suspicious User Profile Identification}

After looking at the inefficiency of Bitly in identifying suspicious links, we proceeded with the detection of suspicious Bitly accounts. On the analysis of all links in encoder's link history, we obtained 112,697 links redirecting to a Bitly warning page (12,344 encoders), giving us more suspicious URLs. To compute the fraction of suspicious links shortened by these encoders, we assigned a \textit{Suspicion Factor (Sus\_Fac)} for each as :
\begin{equation}\label{eq:suspicionFactor}
\small
Sus\_Fac=\frac{\#Links~redirecting~to~warning~page}{\#total~links~collected}
\end{equation}  
We define Sus\_Fac as the ratio of shortened links giving a Bitly warning page to the total links collected for each encoder. Figure~\ref{cumulative} shows the cumulative distribution of number of Bitly users based on their Sus\_Fac. The graph shows that 12,344 users had a Sus\_Fac less than or equal to 1, and 10,326 users had a Sus\_Fac less than or equal to 0.99. This means that 2,018 (12,344 - 10,326) out of 12,344 encoders (16.35\%) had a Sus\_Fac = 1, indicating that they shortened only suspicious links. Also 2,558 encoders (20.72\%) had at least 80\% of their shortened URLs as malicious (Sus\_Fac \textgreater= 0.8). This clearly highlights the malicious intent of these encoders on creating their Bitly accounts. As of now, Bitly follows a \textit{no user suspension} policy and does not even delete a malicious link.~\footnote{\url{http://blog.bitly.com/post/138381844/spam-and-malware-protection}} This facilitates the continued existence of a large number of encoders with such evil motives. All these accounts still exist (as of January 2014) and look legitimate on viewing their profile. Its only when a user visits multiple links from these profiles leading to malicious content, he gets to know that the profile is created for a dedicated purpose of spamming. This approach is quite different from that followed by Twitter, wherein a suspicious user once detected is immediately suspended to prevent further dissemination of malicious content.~\footnote{\url{https://support.twitter.com/articles/18311-the-twitter-rules}} Looking at the extent to which spammers are leveraging the policies used by Bitly, it becomes important for Bitly to also create policies to mitigate this problem. One simple solution could be to assign a credibility score (like Sus\_Fac) to each profile to apprise the users visiting that profile of upcoming risks, if any. This approach has also been explored by Gupta et al. wherein a tweet can be assigned a credibility score based on certain relevant features \cite{22}.  
\subsubsection{Promptness Analysis}
\begin{figure}[h!]
 	\centering
		\includegraphics[scale = 0.52]{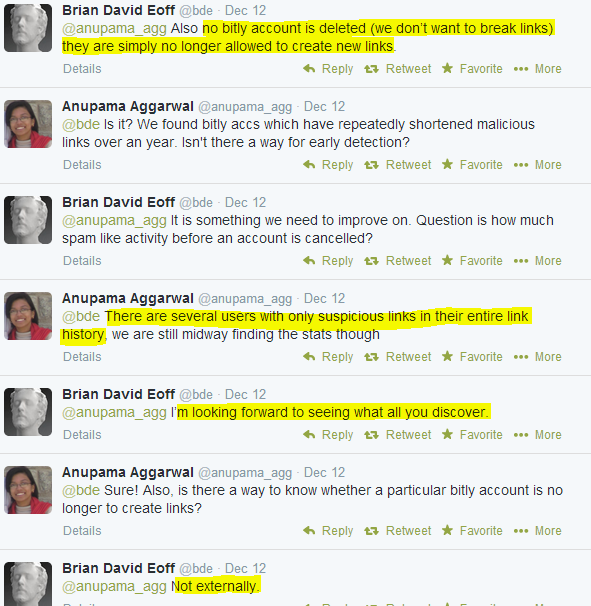}
	\caption{Brian's (Lead Data Scientist at Bitly) reply to out Twitter post about the blog.}
		\label{fig:brianReply}
\end{figure}

Results in the above experiment clearly shows how Bitly users keep shortening only malicious URLs. Till this point of our study, it was unknown to us how Bitly reacts to suspicious user profiles.
 To report our findings and get some insights, we made a blog entry on our initial data analysis.~\footnote{\url{http://precog.iiitd.edu.in/blog/2013/12/bitly-could-do-better/}} Figure~\ref{fig:brianReply} present some comments on our blog by a Lead Data Scientist from Bitly, in which he informed that Bitly does not suspend user accounts but forbids suspicious users from creating any new links. Also, since this information is only known to Bitly and the user, we could not get this data. To verify this claim, we performed another experiment to observe the promptness of Bitly in detecting suspicious profiles. 
 For this, we only considered the highly suspicious profiles obtained from the last experiment (Section~\ref{Efficiency Analysis}). We label a profile as highly suspicious if it has a Sus\_Fac of 1.
 Using this filter, we obtained 80 highly suspicious profiles in our dataset. \textit{encoder\_link\_history} metric from the Bitly API was then used to extract the creation time and number of clicks for all 100 links from each profile. At last, we cumulated the number of links created and clicks received per month and formulated a timeline for each user. The maximum month lag of shortening malicious links that we observed was 24 months for user \textit{bamsesang} (Figure~\ref{fig:bamsesang}), followed by 18 months for user \textit{iplayonlinegames}. The timeline shows that user \textit{bamsesang} shortened all malicious links for 7 months, remained inactive for close to 1 year and then shortened malicious URLs again. On plotting a similar timeline, we found that user \textit{iplayonlinegames} remained active throughout. These users posted links even when the number of clicks received were less. This highlights that they might be posting links randomly and not monitoring their impact. In contrast, some malicious links shortened by these users received a significant number of hits. Out of 80 highly suspicious users that we labeled, 7 users posted only malicious links for more than 5 months. Hence, users shorten only malicious URLs for a prolonged time without getting detected. These results show an extreme delay in suspicious user identification (if at all) by Bitly. 

All this was observed when we only took into consideration past 100 continuous malicious links for each user. There could have been highly active suspicious users who shortened 100 or more malicious links within a single day. Number of months between shortening the first and last suspicious URL by such users could be very high if all their links are studied. Since Bitly API gives only last 100 entries in a single request, this would have required making multiple requests per user to capture the complete link history. We did not do this in our study due to space and time constraints. But in order to check if these highly suspicious users have actually been forbidden by Bitly, we collected their recent link history (after 1 January 2014). We found that 4 of these 80 users were still active and propagating malicious content. 
 Even for the rest, it cannot be said if they have been prohibited from link creation by Bitly or themselves did not create more links. This looks contrary to Bitly's assertion that it disallows suspicious users to shorten more links. This evidently signifies the ease of penetration of spammers on Bitly and delay in its suspicious user detection process which it actually claims to follow.
\subsubsection{Analyzing the effectiveness of Bitly warning page}
After identifying an extreme delay in the detection of suspicious user profiles by Bitly, we inspected if the access to all popular malicious Bitly links eventually die out after Bitly discovers them. Here we define popular malicious Bitly links as the ones with high number of warning pages displayed. This is important to study because it gives a clear picture about the persistence of already identified malicious URL propagation through Bitly network. Also, we restricted our study to only popular links because we wanted to capture the URLs with high overall impact. 

We extracted the top 1,000 Bitly links from our \textit{link-dataset} based on the number of warning pages reported in the month of October 2013. Bitly API was then used to collect the click history of all these links. Next, we determined and separated the links which got recently clicked (after January 2014). Using this measure, we found 352 out of 1,000 malicious links (35.2\%) were also being actively clicked in year 2014. 
This sample study shows that even though Bitly detected these suspicious links months before, users are still getting trapped and visiting these links. These results help us comprehend that a by-passable Bitly warning page alone is not a strong enough measure to curtail the dissemination of spam. Hence, an improved approach could be that Bitly should not only throw a warning page but also block the visit on popular malicious Bitly links already detected. 
\section{Bitly Link Classification}
\vspace{4pt}
In this section, we first describe the feature set used to classify a Bitly URL as malicious or benign. Next, we explain the classification algorithms followed by the experimental setup and results.
\subsection{Feature Selection}\label{Feature Selection for Malicious Bitly URL Detection}
Long URL based features to classify a malicious link has been studied over years. Our target was to inspect if short URL based features also hold some distinctive properties to identify a malicious URL. Since it is difficult to capture the intrinsic characteristics of a landing page by using short URL based features alone, we coupled short URL (Bitly specific) and some long URL based (WHOIS based) features. 
\subsubsection{WHOIS based features}
WHOIS is a query-response protocol that gives information like domain name, domain creation / updation date and domain expiration date for a particular URL. This information is particularly useful to detect domains which are intentionally created for malicious purposes. We used 2 \textit{WHOIS based features}:
\begin{itemize}[leftmargin=0.4cm]
 \item {\bf Domain age}: Most spammers prefer to register their domains for a short duration and also change the domains frequently to evade detection. In Section~\ref{Domain Analysis}, we observed 83.06\% malicious domains to be non-existent when we rechecked them after 5 months. This shows that malicious domains are usually short lived.
 \item {\bf Difference between Link and domain creation}: It is commonly observed that suspicious domains are created / updated just before they are actually used. Hence, we used the difference between domain and Bitly link creation time as one of our feature.
\end{itemize}
\subsubsection{Bitly specific features}
Bitly provides a detailed analytics about each link it shortens. 
 These analytics contain a lot of hidden properties that can help segregate malicious and benign links. We identified some \textit{Bitly specific features} and divided these as \textit{Non-Click} based and \textit{Click} based. \textit{Non-Click} based features define general characteristics of a Bitly URL and are independent of its click history whereas \textit{Click} based features depends on the click analytics of a Bitly link.\\
Three \textit{Non-Click} based features are:
\begin{itemize}[leftmargin=0.4cm]
 \item {\bf Link creation hour}: Malicious users often rely on automated mechanisms to shorten and share the links. Link shortening timestamp patterns in case of such automation might not be similar to the genuine usage trend. Such a behavior can thus be captured by tracking link creation hour of the Bitly links.
 \item {\bf Number of encoders}: Encoders are users who shorten a link using Bitly. Number of encoders corresponding to a Bitly URL depicts its popularity. Malicious communities take advantage of this feature by creating multiple identities to shorten the same link (section~\ref{Connected Network Impact Analysis}). This feature can be used in order to detect the presence of such suspicious communities.
 \item {\bf Type of encoders}: Encoders can either be regular Bitly users or users who use some third party services provided by Bitly. These third party services provide a single interface to shorten and share the links on multiple OSM.~\footnote{\url{https://bitly.com/pages/partners}} Since we collected our data using Twitter, we focused on only Twitter based applications like Twitterfeed~\footnote{\url{http://twitterfeed.com/}}, TweetDeck API~\footnote{\url{https://api.tweetdeck.com/}}, and Tweetbot~\footnote{\url{https://api.tweetdeck.com/}}, etc. 
In addition to these services, various Bitly links were also shortened anonymously by users for which Bitly gives the encoder information as ``someone'' or ``anonymous''. In the \textit{link-metric-dataset}, we observed traces of users who hide their identities and shorten malicious links. Anonymous Bitly link shortening and third party service usage can therefore be used as a feature to identify malicious links.
\end{itemize}
Two \textit{Click} based features are:
\begin{itemize}[leftmargin=0.4cm]
\item {\bf Link creation-click lag}: 
Antoniades et al. stated that most legitimate short URLs are clicked on the same day they are created \cite{3}. The average lifetime of malicious short URLs have also been reported to be higher than that of the legitimate ones \cite{4}. This gives a notion that malicious short URLs do not gain immediate popularity and the number of clicks on such URLs evolve slowly. Hence, we included link creation-click lag as a characteristic feature to capture how quick the short URL resolves.
\item {\bf Type of referring domains}: 
A previous study \cite{12} reveals that large fraction of malicious Bitly URLs get clicked directly through email clients, messengers, chat applications, SMS, etc. Thus, we used the fraction of referring domains that contributes to direct clicks as another feature for our classification. 
\end{itemize}

\subsection{Machine Learning Classification}
\vspace{4pt}
\label{sec:classifiers}
Now we describe the mechanisms used in our classification of malicious short URLs. The experiments involved a 3 step process -- i) creating a labeled dataset, ii) training the suitable machine learning classifier, and iii) testing an unlabeled dataset on the trained classifier. In order to assess the most appropriate and efficient mechanism to detect malicious short URLs, we inspected various machine learning classifiers which were best suited for our study. For this, we used the popular classification algorithms implemented in Weka software package \cite{15}. Weka is an open source collection of machine learning classifiers for data mining tasks. Now, we give a brief description about these classifiers.
\subsubsection{Naive Bayes} It is a simple probabilistic classifier based on the Bayes' theorem. It assumes all classification features to be independent of one another and works best when the dimensionality of inputs is very high. An advantage of using this model is that it does not have a large training data requirement for parameter estimation and classification. It uses variance of variables in each class and is also not sensitive to irrelevant features.
\subsubsection{Decision Tree} It is a popular classification method that uses decision tree as a predictive model. It uses a rule based approach to observe data features and make inferences about item's target value. Decision Tree starts at the root and makes binary (yes / no) decisions at each level until it reaches the leaf node.
\subsubsection{Random Forest} This classifier consists of multiple decision trees and outputs the class which is the mode of the classes output by individual trees. For each data point, it randomly chooses a set of features for classification. It uses averaging to select most important features, hence improve the predictive accuracy and control over-fitting. It can run efficiently on large databases. 
\subsection{Training and Testing Data}
\vspace{4pt}
We divided the \textit{labeled-dataset} into training (75\%) and testing data (25\%). Ten fold cross validation was applied on the training data over all classifiers. Here the data was partitioned into 10 subsets. In each test run, 9 subsets were used for training and 1 was used for testing. This process was repeated 10 times and average accuracy of each classifier was analyzed. The remainder 25\% testing data was then supplied to determine the actual performance of the classifier on an unlabeled dataset.
 
\subsection{Classifier Evaluation}\label{Classifier Evaluation}
\vspace{4pt}
We already defined the feature set and machine learning algorithms used in our classification. This section describes the evaluation metrics and results we obtained from our experiments based on the ground truth dataset.
\subsubsection{Evaluation Metrics}
Two major metrics that we used for the evaluation of our classifier are F-measure (FM) and accuracy (A). F-measure is defined in terms of precision (P) and recall (R). Precision (Equation~\ref{eq:precision}) is the proportion of predicted positives in the class that are actually positive, while recall (Equation~\ref{eq:recall}) is the proportion of the actual positives which are predicted positive. This dependency can be better described by a confusion matrix as presented in Table~\ref{confusionMatrix}. 
\begin{table}[h]
\small
\begin{center}
\begin{tabular}{l|l|c|c|c}
\multicolumn{2}{c}{}&\multicolumn{2}{c}{Predicted Class}&\\
\cline{3-4}
\multicolumn{2}{c|}{}&Malicious&Benign\\
\cline{2-4}
\multirow{2}{*}{Actual Class}& Malicious & $TP$ & $FN$\\
\cline{2-4}
& Benign & $FP$ & $TN$\\
\cline{2-4}
\end{tabular}
\caption{\label{confusionMatrix} Confusion matrix showing relation between True Positives (TP), False Positives (FP), True Negatives (TN), and False Negatives (FN).}
\end{center}
\end{table}
Each row of the confusion matrix represents the instances in an actual class, whereas each column represents the instances in a predicted class. Confusion matrix is very useful in understanding the relation between true positives (correctly identified values), false positives (incorrectly identified values), true negatives (correctly rejected values), and false negatives (incorrectly rejected values). With this knowledge, F-measure and accuracy can be easily computed. F-measure (Equation~\ref{eq:F-measure}) is a weighted average of precision and recall, and accuracy (Equation~\ref{eq:accuracy}) is the closeness of measurements to the actual value. 

\begin{equation}\label{eq:precision}
P = TP/(TP+FP) 
\end{equation}
\vspace{-3ex} 
\begin{equation}\label{eq:recall}
R = TP(TP+FN)
\end{equation}
\vspace{-3ex} 
\begin{equation}\label{eq:F-measure}
FM = 2*(P*R)/(P+R)
\end{equation}
\vspace{-3ex} 
\begin{equation}\label{eq:accuracy}
A = (TP+TN)/(TP+TN+FP+FN)
\end{equation}
\subsubsection{Evaluation Results}
We now describe the results obtained from our classification experiments. We trained and tested 3 classifiers using 7 features described in Section~\ref{Feature Selection for Malicious Bitly URL Detection}. The true positive dataset was split into 75\% training and 25\% testing set. The classifiers were trained on a ground truth of 5,926 malicious and 6,074 benign short URLs and 10 fold cross validation was applied. At last, we evaluated our classifier on the remainder 25\% dataset (2,074 malicious and 1,926 benign Bitly URLs). Table~\ref{classifierAllFeatures} presents the outcome from each of these classifiers. 
\begin{table}[h]
\small
\begin{center}
     \begin{tabular}{|p{2.9cm}|p{1cm}|p{1cm}|p{1cm}|}
    \hline
    {\bf Evaluation Metric}     & {\bf Naive Bayes} & {\bf Decision Tree} & {\bf Random Forest} \\ \hline
    Accuracy              & 72.15\%     & 78.37\%       & \bf{80.43\%}       \\ \hline
    Recall (malicious)    & 73.10\%     & 82.40\%       & 81.00\%       \\ \hline
    Recall (Benign)       & 71.10\%     & 74.10\%       & 79.90\%       \\ \hline
    Precision (malicious) & 73.10\%     & 77.40\%       & 81.20\%       \\ \hline
    Precision (Benign)    & 71.10\%     & 79.60\%       & 79.60\%       \\ \hline
    F-measure (malicious) & 73.10\%     & 76.70\%       & \bf{81.10\%}       \\ \hline
    F-measure (benign)      & 71.10\%     & 76.70\%       & \bf{79.70}\%       \\ \hline
    \end{tabular}
    \caption {\label{classifierAllFeatures} Results of classification of malicious Bitly URLs using all 7 features.}
\end{center}
\end{table}

As we moved from Naive Bayes to Decision Tree, the accuracy and F-measure increased for malicious as well as benign class. With Random Forest, both these metrics increased even further which can be attributed to the point that it considerably reduces false positives in our classification. Hence, we achieved an overall accuracy of 80.43\% and weighted average of F-measure as 80.40\%. 
 We understand that our training dataset is a mix of Bitly links that receive and do not receive clicks. Out of the 8,000 malicious links in our \textit{labeled-dataset}, we found that 3,693 (46.16\%) links were never clicked. Although this property itself serves as a feature in the classification, we believe that our classifier should perform even better if we segregate all links with zero clicks. This is because such links can be easily classified using only \textit{Non-Click} based features. Hence, we separated Bitly links with zero clicks and obtained a true positive malicious dataset of 3,693 from a total of 8,000 links. We also randomly selected 3,693 benign links with no clicks from the \textit{labeled-dataset} and performed our classification experiments again by removing \textit{Click} based features. Table~\ref{classifierNonClickFeatures} gives the results from this classification. As expected, an increase in the overall accuracy and F-measure for each classifier was observed. With Random Forest, we achieved an accuracy of 86.41\% and weighted average of F-measure as 86.40\%. When the complete \textit{labeled-dataset} was tested again by excluding \textit{Click} based features, maximum accuracy reached 83.50\% and F-measure for malicious class increased to 84.1\%.

Previous studies on phishing also shows that Random Forest works the best as compared to other classifiers \cite{9,17}. In order to further inspect the performance of Random Forest, we analyzed the confusion matrix for complete as well as \textit{Non-Click} \textit{labeled-dataset} (Table~\ref{fig:confusionMatrixAllMain}). For complete \textit{labeled-dataset}, we could correctly identify 81\% of malicious short URLs but 19\% were misclassified as benign. On a manual inspection of some of these misclassified profiles, we observed their click pattern to be quite similar to that of the benign profiles. In case of \textit{Non-Click} \textit{labeled-dataset}, the correct classification rate reached 89.60\% and misclassification rate dropped to almost half (10.40\%). Also, on removing \textit{Click} based features from complete \textit{labeled-dataset}, true positive increased to 84.20\% and false negative decreased to 15.8\%. This indicated that our \textit{Non-Click} based features are alone good at classifying a malicious Bitly link, irrespective of looking at its click history. Although our proposal is capable of capturing the click patterns of a link, it can be particularly helpful to detect malicious Bitly links even before they are clicked.
\begin{table}[h]
\small
\begin{center}
     \begin{tabular}{|p{2.9cm}|p{1cm}|p{1cm}|p{1cm}|}
    \hline
    {\bf Evaluation Metric}     & {\bf Naive Bayes} & {\bf Decision Tree} & {\bf Random Forest} \\ \hline
    Accuracy              & 80.02\%     & 85.06\%       & {\bf 86.41}\%       \\ \hline
    Recall (malicious)    & 79.60\%     & 89.50\%       & 89.60\%       \\ \hline
    Recall (Benign)       & 80.40\%     & 80.80\%       & 83.40\%       \\ \hline
    Precision (malicious) & 79.30\%     & 81.50\%       & 83.60\%       \\ \hline
    Precision (Benign)    & 80.70\%     & 89.10\%       & 89.50\%       \\ \hline
    F-measure (malicious) & 79.50\%     & 85.30\%       & 86.50\%       \\ \hline
    F-measure (benign)    & 80.50\%     & 84.80\%       & {\bf 86.30}\%       \\ \hline
    \end{tabular}
    \caption {\label{classifierNonClickFeatures} Results of classification of malicious Bitly URLs using only \textit{Non-Click} based features.}
\end{center}
\end{table}
\subsubsection{Feature Ranks}
Some features in machine learning classification are more important than others. Now we describe the most informative features based on their ranks using Weka's \textit{InfoGainAttributeEval} package for attribute selection.~\footnote{\url{http://weka.sourceforge.net/doc.dev/weka/attributeSelection/InfoGainAttributeEval.html}} This technique evaluates the worth of an attribute by measuring the information gain with respect to the class. Table~\ref{tb:featureRank} gives the ranked feature lists for classification experiments on complete dataset. 
\begin{table}[ht]
\small
\centering
\begin{tabular}{|l|l|}\hline
\small
\textbf{Rank} & \textbf{Feature}\\ \hline
1 & Type of referring domains\\ \hline
2 & Difference between Link and domain creation\\ \hline
3 & Domain age\\ \hline
4 & Link creation hour\\ \hline
5 & Type of encoders \\ \hline
6 & Link creation-click lag \\ \hline
7 & Number of encoders \\ \hline
\end{tabular}
\caption{\label{tb:featureRank} Feature rank based on information gain for classification on complete \textit{labeled-dataset}.}
\end{table}
On a mix dataset of clicked and non-clicked links, \textit{Type of referring domains} is found to be the most discriminating feature. Malicious Bitly links usually have more direct referrers, i.e. such links are propagated more through email clients, mobile applications, SMS, etc. Such sources are often targeted by spammers because they have a direct impact on the user, which increases the chances for a malicious link to be seen and visited. \textit{Difference between Link and domain creation} is another important feature based on information gain. As seen in Section~\ref{Domain Analysis}, offenders buy cheap domains for the deliberate purpose of spamming. Such domains are mostly registered for a short time span with the objective to fully exploit them before they expire. The third most informative feature \textit{Domain age} highlights the same. 

Another strategy adopted by spammers is to shorten links at odd hours of the day, as depicted by the fourth most informative feature \textit{Link creation hour}. \textit{Type of encoders}, \textit{Link creation-click lag}, and \textit{Number of encoders} are found to be comparatively less informative and did not help much in the classification. For our Non-Click dataset, \textit{Link creation hour} is observed to be the most important feature. Remaining features follow the same ranking pattern as in the complete dataset. 
\begin{table*}[ht]
\small
\begin{center}
\begin{tabular}{l|l|c|c|c}
\cline{3-4}
\multicolumn{2}{c|}{}&Malicious&Benign\\
\cline{2-4}
& Malicious & $81.00\%$ & $19.00\%$\\
\cline{2-4}
& Benign & $20.10\%$ & $79.90\%$\\
\cline{2-4}
\end{tabular}
\quad
\begin{tabular}{l|l|c|c|c}
\cline{3-4}
\multicolumn{2}{c|}{}&Malicious&Benign\\
\cline{2-4}
& Malicious & $89.60\%$ & $10.40\%$\\
\cline{2-4}
& Benign & $16.60\%$ & $83.40\%$\\
\cline{2-4}
\end{tabular}
\quad
\begin{tabular}{l|l|c|c|c}
\cline{3-4}
\multicolumn{2}{c|}{}&Malicious&Benign\\
\cline{2-4}
& Malicious & $84.20\%$ & $15.80\%$\\
\cline{2-4}
& Benign & $17.20\%$ & $82.80\%$\\
\cline{2-4}
\end{tabular}
\caption{\label{fig:confusionMatrixAllMain} Confusion matrix for - (a) complete \textit{labeled-dataset}; (b) only Non-Click data from \textit{labeled-dataset}; (c) complete \textit{labeled-dataset} by applying only \textit{Non-Click} based features.}
\end{center}
\end{table*} 
\section{Conclusion}\label{Conclusion}
\vspace{4pt}
Our initial analysis gave an overview of some characteristics of malicious Bitly links and their propagation on OSM. Spammers buy cheap domains for a dedicated purpose of spamming, which eventually die out after targeting a significant number of victims. 
We found that spammers exploit Bitly's policy of not imposing a cap on the number of connected Facebook / Twitter accounts. We traced such spammers and detected 3 malicious communities which operate across Bitly and Twitter. 
 We then unveiled some loopholes in Bitly's security policies. Bitly could not effectively detect malicious URLs already tracked by the popular blacklist services like APWG and VirusTotal. Leaving this aside, we also observed a lack of effectiveness in using the claimed detection services (like SURBL) by Bitly. Malicious users are found to abuse Bitly's \textit{no account suspension} policy. About sixteen percent malicious users in our dataset shortened only suspicious links in their history without being noticed. Bitly asserts that it forbids such users from shortening more links, but we identified users who kept shortening malicious links for close to 2 years and their state of detection is still unknown. We also highlighted how a by-passable Bitly warning page is only partially effective to curtail the problem of spam.

At last, we proposed a mechanism to detect malicious links on Bitly. We identified features from Bitly and coupled them with some domain specific features to classify a Bitly URL as malicious or benign. Computation of these features required close to 50 seconds per Bitly URL. Our classifier predicted malicious links with the best accuracy of 80.40\% on a mix dataset of clicked and non-clicked links. By eliminating \textit{Click} based features, our classifier attained an improved accuracy of 83.51\% on the mix dataset, and 86.41\% on the \textit{Non-Click} dataset. Thus, our algorithm is not only efficient in detecting malicious Bitly links when they receive clicks, but can also identify them much before they target the netizens. This shows that in addition to the blacklists and other spam detection filters, Bitly specific feature set can also be used to detect malicious content.
\section{Limitations and Future Work}
\vspace{4pt}
Even though our dataset is unique, it is a limited dataset acquired from Bitly. This dataset captured only links detected malicious by Bitly in the month of October 2013. Since the characteristics of spammers evolve over time, we would like to do a detailed comparative analysis on more exhaustive dataset. Bitly claims to forbid certain suspicious accounts but does not give this information through its API. This created a trouble in interpreting if a dormant suspicious account in our dataset has been blocked by Bitly or not shortening the links itself. 
 For malicious Bitly link detection, our classifier currently works better on \textit{Non-Click} dataset. This is because we used only 2 \textit{Click} based features in our classification and links with zero clicks did not have click information. In future, we would like to do a temporal study of how click patterns on a malicious Bitly URL evolve over time, which can serve as another good feature for our classifier. Our feature set can be broadened and generalized to detect spam from any short URL services. In addition, we would like to develop a browser extension that can work in real time and classify any short link as malicious or benign. Such a service can be very useful in short URL spam detection across multiple online media.
\section*{Acknowledgment}
\vspace{4pt}
We thank all the members of Precog research group at IIIT-Delhi for their valuable feedback and support throughout this work. We would also like to express our sincere gratitude to Bitly and particularly Brian David Eoff (senior data scientist) and Mark Josephson (CEO) for sharing the data with us. We also thank CERC at IIIT-Delhi for their encouragement and insightful comments.



\begin{thebibliography} {9}
\bibitem{1} A. Neumann, J. Barnickel, U. Meyer. Security and Privacy Implications of URL Shortening Services. In proceedings of Web 2.0 Security and Privacy (W2SP) (2011).
\bibitem{2} F. Klien, M. Strohmaier. Short Links Under Attack: Geographical Analysis of Spam in a URL Shortener Network. In proceedings of the 23rd ACM conference on Hypertext and social media (2012), Pages 83-88. 
\bibitem{3} D. Antoniades, I. Polakis, G. Kontaxis. we.b: The web of short URLs. In proceedings of the 20th international conference on World wide web (2011), Pages 715-724. 
\bibitem{4} F. Maggi, A. Frossi, S. Zanero, G. Stringhini, B. Stone-Gross, C. Kruegel, G. Vigna. Two Years of Short URLs Internet Measurement: Security Threats and Countermeasures. In proceedings of the 22nd international conference on World Wide Web (2013), Pages 861-872. 
\bibitem{5} V. Kandylas and A. Dasdan. The Utility of Tweeted URLs for Web Search. In proceedings of the 19th international conference on World wide web (2010), Pages 1127-1128.  
\bibitem{6} K. Thomas, C. Grier, J. Ma, V. Paxson, and D. Song. Design and Evaluation of a Real-Time URL Spam Filtering Service. Security and Privacy (SP) IEEE Symposium (2011), Pages 447 - 462.
\bibitem{7} H. Gao, J. Hu, and C. Wilson. Detecting and Characterizing Social Spam Campaigns. In proceedings of the 10th ACM SIGCOMM conference on Internet measurement (2010), Pages 35-47.
\bibitem{8} F. Benevenuto, G. Magno, T. Rodrigues, and V. Almeida. Detecting Spammers on Twitter. In Collaboration, Electronic messaging, Anti-Abuse and Spam Conference (CEAS) (2010).
\bibitem{9} A. Aggarwal, A. Rajadesingan, and P. Kumaraguru. PhishAri: Automatic Realtime Phishing Detection on Twitter.  In Seventh IEEE APWG eCrime researchers summit (eCRS) (2012). Master's thesis, IIIT-Delhi (2012).
\bibitem{10} C. Grier, K. Thomas, V. Paxson, and M. Zhang. @spam: The Underground on 140 Characters or Less. In proceedings of the 17th ACM conference on Computer and communications security (2010), Pages 27-37.
\bibitem{11} S. Lee and J. Kim. WARNINGBIRD: Detecting Suspicious URLs in Twitter Stream. NDSS 2012 (2012).
\bibitem{12} D. Wang, S. B. Navathe, L. Liu, D. Irani, A. Tamersoy, and C. Pu. Click Traffic Analysis of Short URL Spam on Twitter. Collaborative Computing: Networking, Applications and Worksharing (Collaboratecom), 2013 9th International Conference (2013), Pages 250-259.
\bibitem{13} S. Chhabra, A. Aggarwal, F. Benevenuto, and P. Kumaraguru. Phi.sh/\$oCiaL: The Phishing Landscape through Short URLs. CEAS '11 Proceedings of the 8th Annual Collaboration, Electronic messaging, Anti-Abuse and Spam Conference (2011), Pages 92-101.
\bibitem{14} S. Yoon, J. Park, C. Choi, and S. Kim. Poster : SHRT – New method of URL shortening including relative word of target URL. In symposium on Usable Privacy and Security (SOUPS) (2013).
\bibitem{15} M. Hall, E. Frank, G. Holmes, B. Pfahringer, P. Reutemann, and I. H. Witten. The WEKA Data Mining Software: An Update. ACM SIGKDD Explorations Newsletter (2009), vol. 11, no. 1, Pages 10–18.
\bibitem{16} Symantec. Snapchat Spam: Sexy Photos Lead to Compromised Branded Short Domains. \url{http://www.symantec.com/connect/blogs/snapchat-spam-sexy-photos-lead-compromised-branded-short-domains}, January 2014.
\bibitem{17} S. Abu-Nimeh, D. Nappa, X. Wang, and S. Nair. A Comparison of Machine Learning Techniques for Phishing Detection. In proceedings of eCrime researchers summit (2007), ACM, Pages 60-69.
\bibitem{18} Mashable. Warning: Bit.ly Is Eating Other URL Shorteners for Breakfast. \url{http://mashable.com/2009/10/12/bitly-domination/}, October 2009.
\bibitem{19} Symantec. Spam with .gov URLs. \url{http://www.symantec.com/connect/blogs/spam-gov-urls}, October 2012.
\bibitem{20} Symantec. Malicious Shortened URLS on Social Networking Sites. \url{http://www.symantec.com/threatreport/topic.jsp?id=threat\_activity\_trends\&aid=malicious\_shortened\_urls}, 2010.
\bibitem{21} theguardian. Facebook spammers make \$200m just posting links, researchers say. \url{http://www.theguardian.com/technology/2013/aug/28/facebook-spam-202-million-italian-research}, 2013.
\bibitem{22} A. Gupta and P. Kumaraguru. Credibility ranking of tweets during high impact events. In Proceedings of the 1st Workshop on Privacy and Security in Online Social Media (PSOSM), In conjunction with WWW'12 (2012). 
\bibitem{23} Google safebrowsing. \url{https://developers.google.com/safe-browsing/}
\bibitem{24} SURBL. \url{http://www.surbl.org/}
\bibitem{25} URL shortening service, Bitly. \url{https://bitly.com/}
\bibitem{26} Bitly, spam and malware protection. \url{http://blog.bitly.com/post/138381844/spam-and-malware-protection}
\bibitem{27} About Bitly. \url{http://www.enterprise.bitly.com/about-us/}
\bibitem{28} Branded short domains on Bitly. \url{https://bitly.com/a/features}
\bibitem{29} URL shortening service, goo.gl. \url{https://goo.gl/}
\end{thebibliography}
%


\end{document}